\documentclass[]{aa}

\usepackage{txfonts}
\usepackage{graphicx}
\usepackage{epsfig}
\usepackage{natbib}
\usepackage{ulem}

\newcommand{\hda}{HD101412 }
\newcommand{\hdb}{HD135344 B }
\newcommand{\hdc}{HD179218 }
\bibpunct{(}{)}{;}{a}{}{,} % to follow the A&A style

\begin{document}

\title{The Structure of Protoplanetary Disks Surrounding Three Young Intermediate Mass Stars \thanks{Based on observations collected at the European Southern Observatory, Paranal, Chile. (Program ID 077.C-0521A)}}

\subtitle{I. Resolving the disk rotation in the [OI] 6300 \AA\, line}

\author{G. van der Plas\inst{1,2}
  \and M. E. van den Ancker\inst{1}
  \and D. Fedele\inst{1,3,4}
  \and B. Acke\inst{5}
  \and C. Dominik\inst{2}
  \and L.B.F.M. Waters\inst{2}
  \and J. Bouwman\inst{4}}

\offprints{G. van der Plas, \email{gvanderp@eso.org}}

\institute{European Southern Observatory, Karl-Schwarzschild-Str.2, D 85748 Garching bei M\"unchen, Germany
  \and Sterrenkundig Instituut 'Anton Pannekoek', University of Amsterdam, Kruislaan 403, 1098 SJ Amsterdam, The Netherlands
  \and Dipartimento di Astronomia, Universit\'a Degli Studi Di Padova, Vicolo Dell\`\, Osservatorio 2, 35122 Padova, Italy
  \and Max-Planck-Institut f\"ur Astronomie, K\"onigstuhl 17, 69117 Heidelberg, Germany 
  \and Institute of Astronomy, KU Leuven, Celestijnenlaan 200D, 3001 Leuven, Belgium\fnmsep\thanks{Postdoctoral Fellow of the Fund for Scientific Research, Flanders.}}

%\date{}

\abstract{

We present high spectral resolution optical spectra of three young intermediate mass stars, in all of which we spectrally resolve the  6300 \AA\, [OI] emission line. Two of these have a double peaked line profile. We fit these data with a simple model of the [OI] emission caused by photo-dissociation of OH molecules in the upper layer of a circumstellar disk by stellar UV radiation and thus translate the Doppler broadened [OI] emission profile into an amount of emission as a function of distance from the central star. The resulting spectra are in agreement with the expected disk shapes as derived from their spectral energy distribution. We find evidence for shadowing by an inner rim in the disk surrounding \hda and see a flaring disk structure in \hdc while the [OI] spectrum of \hdb is more complex. The [OI] emission starts for all three targets at velocities corresponding to their dust sublimation radius and extends up to radii of 10 -- 90 AU. This shows that this method can be a valuable tool in the future investigation of circumstellar disks.

}

\maketitle

\section{Introduction}

HD101412, \hdb and \hdc are three isolated young intermediate mass stars. Two of them, \hda and \hdc are Herbig Ae/Be (HAEBE) stars while the third, HD135344 B (often incorrectly referred to as HD135344 in the literature), is a classical T Tauri star of spectral type F. HAEBE stars are the more massive counterparts of T Tauri pre-main-sequence stars with typical masses of 2 - 8 M$_\odot$. Their spectral energy distribution (SED) is characterized by the presence of an infrared excess due to thermal re-emission of circumstellar dust which is thought to be the signature of a circumstellar disk. Previous studies of these stars and their circumstellar disks have focused mainly on this dust (e.g \cite{1992ApJ...397..613H}, \cite{2001A&A...371..186N}, \cite{1997ApJ...490..792M} and \cite{2001A&A...365..476M}).

 Dust however contains only  1 \% of the mass of the disk and it is unclear whether this dust is coupled to  the other 99\% of the disk mass, the gas. To check this coupling and to get a better picture of circumstellar disks in general, it is necessary to study the dust as well as the gas located in circumstellar disks simultaniously. Next to the abundance, there are other reasons such as its potential to probe the dynamics and the physical and chemical structure of disks that illustrate why the circumstellar gas is of interest. Historically it has been very difficult to study the circumstellar gas, but the development of bigger mirrors and better spectrographs in the recent years has opened the way for several gas diagnostics. These include hot and cold CO gas (e.g. \cite{2004A&A...427L..13B}; \cite{2004ApJ...617.1167B}; \cite{2004ApJ...606L..73B} and \cite{2003ApJ...589..931N}), hot water (e.g. \cite{2004ApJ...603..213C}) and molecular hydrogen (e.g. \cite{2008A&A...477..839C}; \cite{2007ApJ...661L..69B} and \cite{2007ApJ...666L.117M}). 

 \cite{2005A&A...436..209A} and \cite{2006A&A...449..267A} have added molecular oxygen to this list. They have developed a method for examining excited neutral oxygen atoms in the surface of a flared, rotating passive disk that they have tested on 2 HAEBE stars: HD97048 and HD100546. With their method, they resolve the disk rotation and the distribution of the emitting gas in the surface layer of the circumstellar disk and also find evidence for a gap in the disk around HD100546 that they interpret as due to the presence of an orbiting body of planetary mass within the circumstellar disk.

 In this paper we apply the method of \cite{2005A&A...436..209A} on three more stars, one of which has a non-flaring (flat) disk geometry. We first describe these stars, together with  our observations and data reduction in Sections \ref{sec:obs} and \ref{sec:stel_par}, then proceed to describe the method and the resulting spectra in Section \ref{sec:analysis} and discuss these results in Section \ref{sec:dis_con}.
 We have also observed our targets with VLT/MIDI to get spatially resolved observations of the 10 $\mu $m silicate bump whose emission originates from warm dust in the same region as the [OI] emission. The results of these interferometric observations will be presented in Fedele et al. (2008, in prep; Paper II). 

\section{Observations and Data Reduction}
\label{sec:obs}

We obtained high spectral resolution \'echelle spectra of \hda, \hdb and \hdc during 4 nights with the KUEYEN 8.2 meter ESO VLT coupled to the UVES spectrograph \citep{2000SPIE.4008..534D}.  The observation log is presented in table \ref{table:dataset}. The data was reduced with the UVES\footnote{http://www.eso.org/instruments/uves/} pipeline\footnote{http://www.eso.org/projects/dfs/dfs-shared/web/uves/uves-pipe-recipes.html} which performs wavelength calibration, order extraction, background subtraction and flatfield correction.
The observations consist of 4 data sets per target. Half of these sets were taken with the slit unrotated (0 degrees), and the other half with the slit rotated 90 degrees. The spectra have a spectral resolution of $\frac{\lambda}{\Delta\lambda} = 77000$, determined using the telluric absorption lines around the 6300 \AA\,  [OI] feature. All three sources are spatially unresolved. We derive upper limits on their angular size  using a 3 $\sigma$ detection limit and following the same procedure as e.g.  \cite{1994A&A...285..929H}, \cite{1997A&AS..126..437H} and \cite{2006A&A...449..267A}. The  {\it upper limits} to the size of the emitting region are 0.25 arcsec corresponding to 39 AU for HD101412, 0.26 arcsec or 36 AU for \hdb and 0.22 arcsec or 53 AU for \hdc. These sizes represent the radii that contain 95$\%$ of the flux in the 6300 \AA\, [OI] line. The observations were made at two different dates to coincide with VTLI/MIDI observations of the same targets in the mid-infrared (N-band). Since the spectra taken with different slit orientation show negligible variation they have been averaged. The data procured at different times however do show temporal change and therefore are handled independently. They  will be referred to as B1 (May 15 - 17) and B2 (June 16 \& July 3, 4) in this paper. 
\medskip

\begin{table}
\setlength{\tabcolsep}{1.1mm}
\caption{Log of spectroscopic observations with UVES}              % title of Table
\label{table:dataset}      % is used to refer this table in the text
\centering                                      % used for centering table
\begin{tabular}{c c c c c c c}          % centered columns (7 columns)
\hline\hline                        % inserts double horizontal lines
Object & Date & UT & Exp. time & PA & B & S/N\\% table heading
& dd/mm/yy & hh:mm:ss & [s] & [$^{\circ}$] & & \\    % table heading
\hline
                                   % inserts single horizontal line
  \hda   & 15/05/06 & 23:37:48 & 300 & 0  & B1 & 174\\    
         & 15/05/06 & 23:43:45 & 300 & 0  & B1 & 199\\    
         & 15/05/06 & 23:50:04 & 300 & 0  & B1 & 189\\    
         & 15/05/06 & 23:56:01 & 300 & 0  & B1 & 175\\    
         & 17/05/06 & 00:39:17 & 300 & 90 & B1 & 183\\    
         & 17/05/06 & 00:45:15 & 300 & 90 & B1 & 207\\    
         & 17/05/06 & 00:51:13 & 300 & 90 & B1 & 155\\    
         & 17/05/06 & 00:57:09 & 300 & 90 & B1 & 199\\    
         & 03/07/06 & 23:52:01 & 300 & 0  & B2 & 149\\    
         & 03/07/06 & 23:58:00 & 300 & 0  & B2 & 141\\   
         & 04/07/06 & 00:03:58 & 300 & 0  & B2 & 161\\   
         & 04/07/06 & 00:09:53 & 300 & 0  & B2 & 151\\   
         & 04/07/06 & 00:18:58 & 300 & 90 & B2 & 137\\   
         & 04/07/06 & 00:24:53 & 300 & 90 & B2 & 144\\   
         & 04/07/06 & 00:30:48 & 300 & 90 & B2 & 135\\   
         & 04/07/06 & 00:36:46 & 300 & 90 & B2 & 155\\   

\hline   
    \hdb & 17/05/06 & 04:19:43 & 300 & 0  & B1 & 197\\
         & 17/05/06 & 04:25:39 & 300 & 0  & B1 & 196\\
         & 17/05/06 & 04:35:08 & 300 & 90 & B1 & 234\\
         & 17/05/06 & 04:41:04 & 300 & 90 & B1 & 262\\	       				  
         & 16/06/06 & 03:49:40 & 300 & 0  & B2 & 144\\
         & 16/06/06 & 03:55:36 & 300 & 0  & B2 & 143\\
         & 16/06/06 & 04:06:12 & 300 & 90 & B2 & 227\\
         & 16/06/06 & 04:12:29 & 300 & 90 & B2 & 240\\
     
\hline     
    \hdc & 17/05/06 & 06:50:00 & 180 & 0  & B1 & 309\\
         & 17/05/06 & 06:56:28 & 180 & 90 & B1 & 237\\  		       				  
         & 16/06/06 & 04:22:43 & 180 & 0  & B2 & 293\\
         & 16/06/06 & 04:28:05 & 180 & 90 & B2 & 251\\

\hline                                             %inserts single line
\end{tabular}
\end{table}

\begin{figure}
  \resizebox{\hsize}{!}{\includegraphics{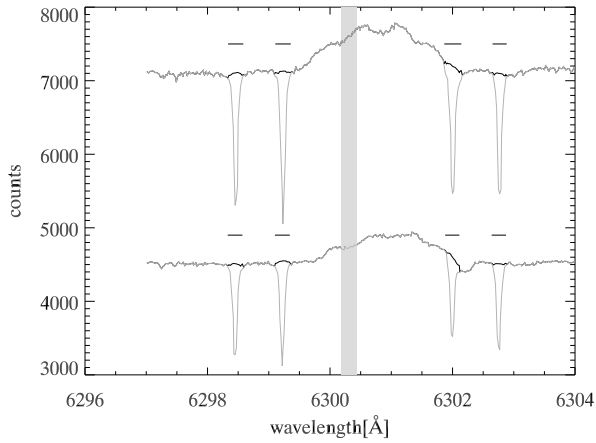}}
  \caption{The spectrum for HD101412. The clipped region is shown in black for the B1 (upper line) and the B2 (lower line) spectra, and is also shown with horizontal bars above the clipped parts. The grey area denotes the location of the (subtracted) 6300 \AA\, [OI] sky line. The emission is not symmetric around this line because these spectra are not yet corrected for the radial and barycentric velocity %(\textit{so, should i show the velocity corrected spectra? if so, they overlap with fig. 4 and 5. combine these?})
}
  \label{fig:telluric_correction_101412}
\end{figure}

\begin{figure}
  \resizebox{\hsize}{!}{\includegraphics{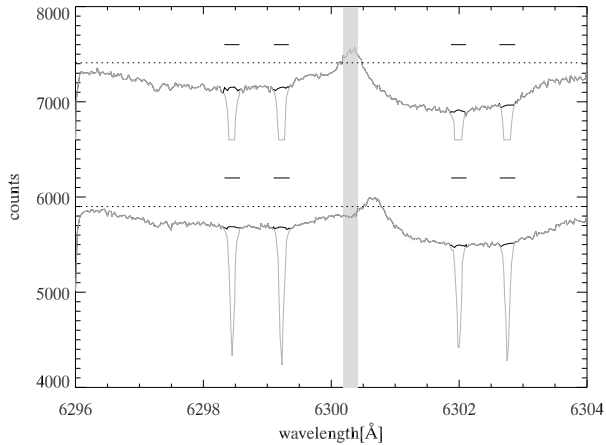}}
  \caption{Same as Figure~\ref{fig:telluric_correction_101412} for \hdb. Note that the   6300 \AA\, [OI] line is mixed with broad atmospheric absorption lines. The over plotted horizontal dotted line indicates the continuum and the telluric absorption lines of the B1 data have been truncated for so it wont overlap with the B2 data. }
  \label{fig:telluric_correction_135344}
\end{figure}

\begin{figure}
  \resizebox{\hsize}{!}{\includegraphics{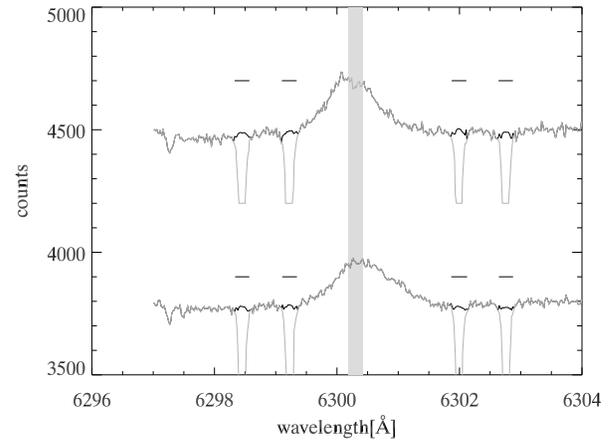}}
  \caption{Same as Figure~\ref{fig:telluric_correction_101412} for \hdc , the telluric absorption lines of the B1 data have been truncated for so they wont overlap with the B2 data.}
  \label{fig:telluric_correction_179281}
\end{figure}

 We determined the radial velocity for our targets from the difference between the measured line center and the theoretical line center for 9 to 31 atmospheric absorption lines observed in the same spectrum.  The errors have been determined by bootstrapping the data set, i.e. by estimating the variability of the radial velocity from the variability of this velocity between sub samples. The size of the sub samples used are 90\% of the total sample and we have estimated the variability by calculating the mean of the subsample a 1000 times. The results are shown in Table~\ref{table:stellar_parameters}. The [OI] 6300 \AA\, line is located in between several telluric absorption lines. To remove these, we fit and fill the absorption lines with a Gaussian, and average the telluric line region over 3/2 FWHM to suppress telluric line residuals.  These telluric absorption lines only overlap the [OI] spectral region on the red wing of HD101412. The results are shown in Figures \ref{fig:telluric_correction_101412} -~\ref{fig:telluric_correction_179281}. In these Figures we also show the location of the 6300 \AA\, [OI] sky emission line. This line has been removed by calculating the sky emission from averaging above and below the echelle order and subtracting this from the final spectrum. Because the sky emission is very weak compared to the flux collected from the targets, the procedure has no significant effect on either the resulting spectrum or the signal to noise.
 
%\begin{figure}
%  \resizebox{\hsize}{!}{\includegraphics{radial_velocity_variation_HD101412.eps}}
%  \caption{Radial velocity determination for \hda from 40 lines. The values for the radial velocity have a mean value of  16.9 $\pm$ 0.2 km s$^{-1}$ (grey vertical line)}
%  \label{fig:radial_velocity_variation_HD101412}
%\end{figure}%

%\begin{figure}
%  \resizebox{\hsize}{!}{\includegraphics{radial_velocity_variation_HD179218.eps}}
%  \caption{Radial velocity determination for \hdc from 18 lines. The values for the radial velocity have a mean value of 15.4 $\pm$ 2.3 km s$^{-1}$  }
%  \label{fig:radial_velocity_variation_HD179218}
%\end{figure}

\begin{figure}
%  \resizebox{\hsize}{!}{\includegraphics{velocityplot_101412_difference.eps}}
 \resizebox{\hsize}{!}{\includegraphics{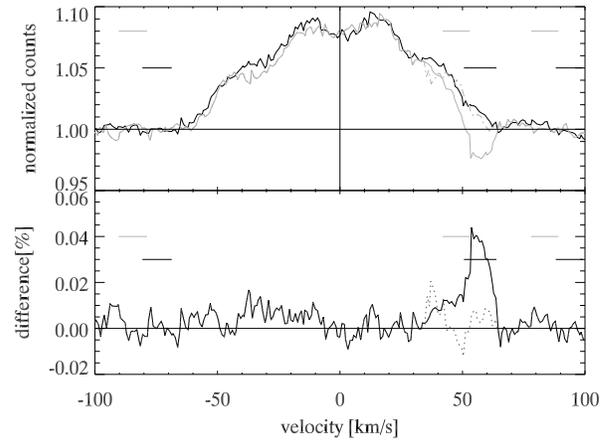}}
% \caption{In the top window the velocity spectrum of \hda centered around the [OI] line, after correcting for barycentric and radial velocity and normalization, for B1(black) and B2(grey).  The difference between both observations is shown in the bottom window. The interstellar absorption feature at 50 km s$^{-1}$ discussed in section \ref{sec:obs} is shown with a dotted line and replaced with the blue shifted part of the same spectrum.}
  \caption{In the top window the velocity spectrum of \hda centered around the [OI] line, after correcting for barycentric and radial velocity and normalization, for B1(black) and B2(grey).  The location of the clipped out telluric lines is shown by horizontal bars. The difference between both observations is shown in the bottom window. The interstellar absorption feature at 50 km s$^{-1}$ discussed in section \ref{sec:obs} is clearly visible and replaced with the blue shifted part of the same spectrum as shown with the dotted line.}
  \label{fig:velocity_correction_hda_difference}
\end{figure}

\begin{figure}
  \resizebox{\hsize}{!}{\includegraphics{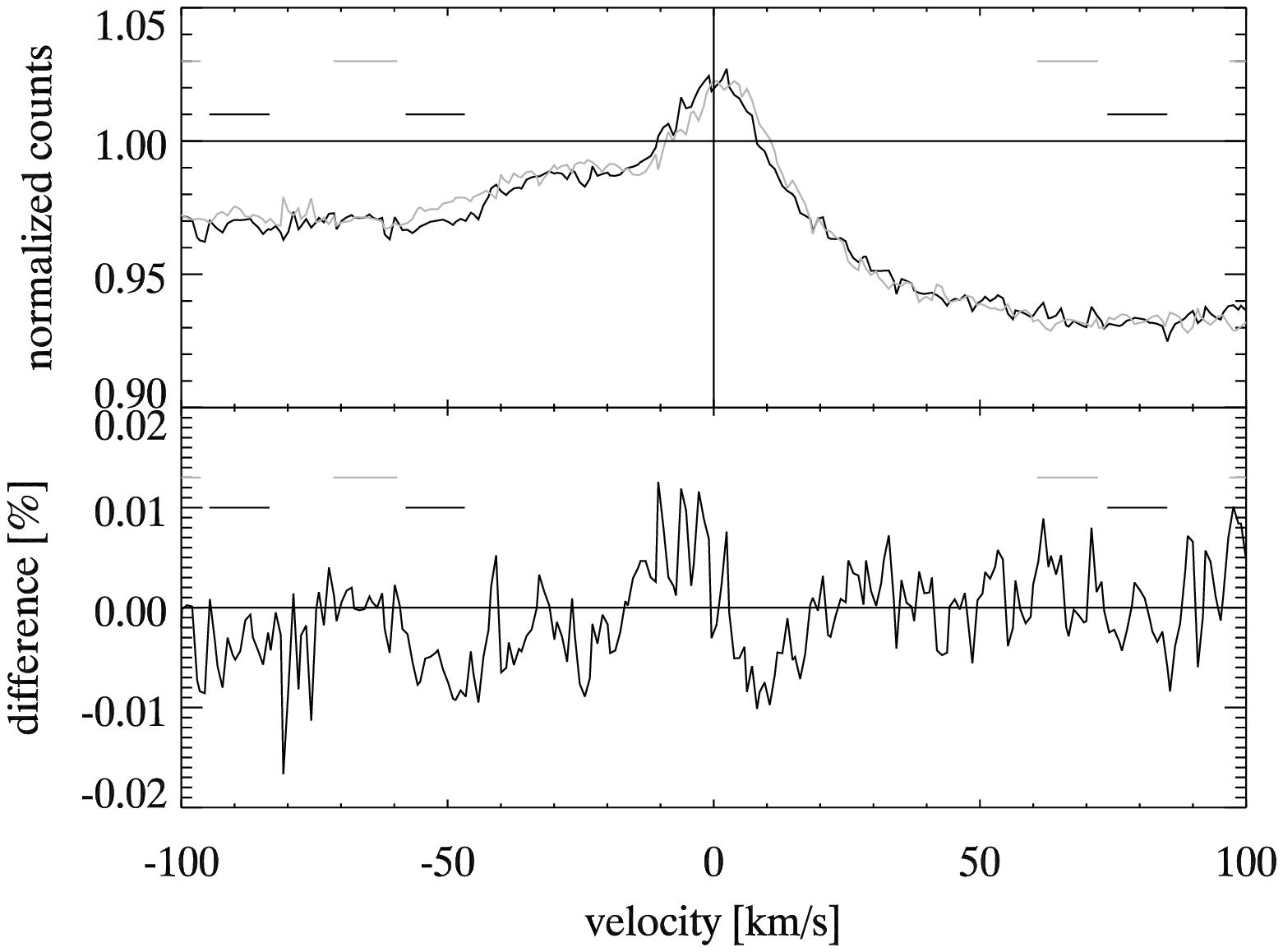}}
  \caption{Same as Figure \ref{fig:velocity_correction_hda_difference} for \hdb}
  \label{fig:velocity_correction_hdb_difference}
\end{figure}

\begin{figure}
  \resizebox{\hsize}{!}{\includegraphics{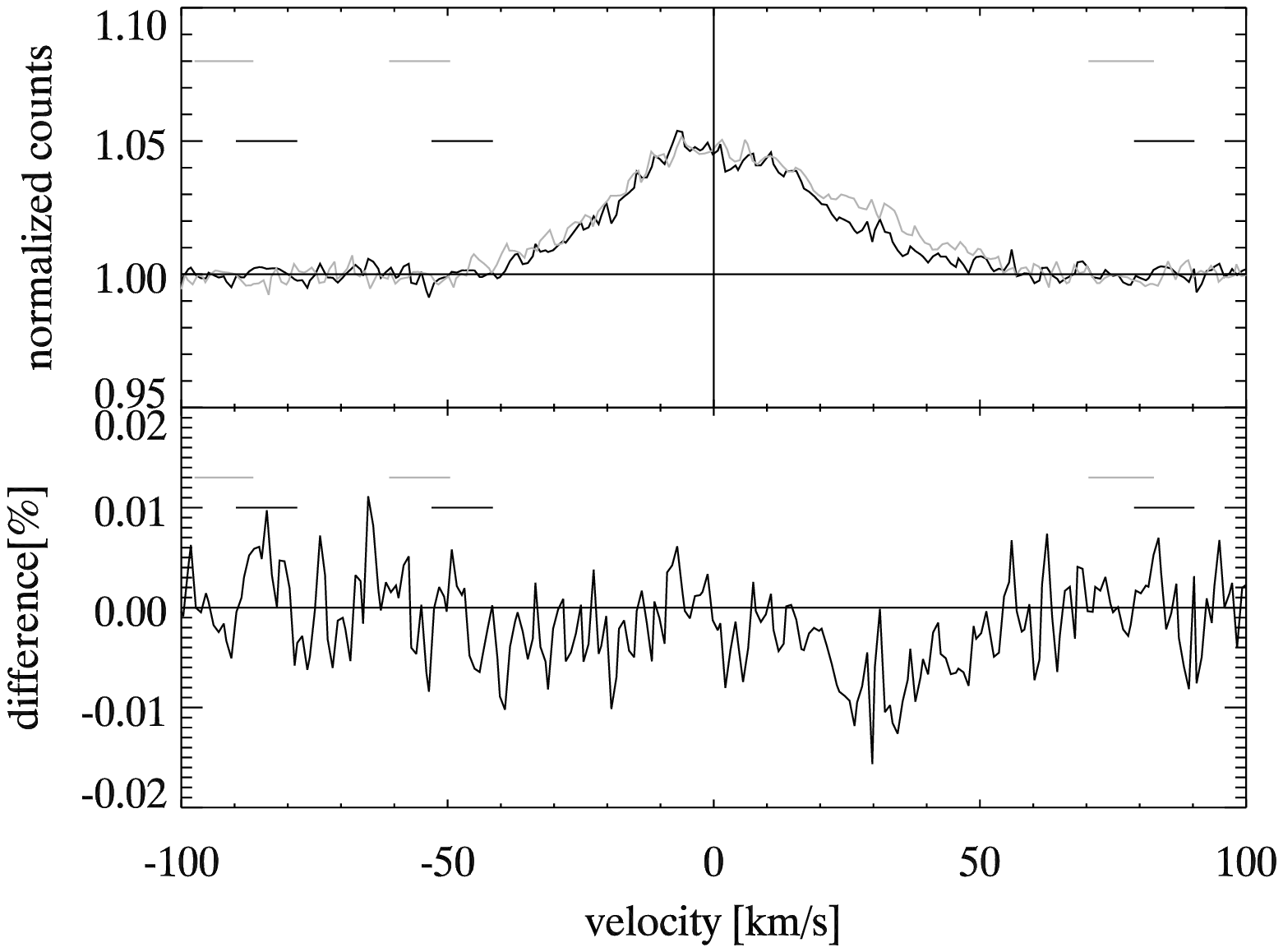}}
  \caption{Same as Figure \ref{fig:velocity_correction_hda_difference} for \hdc}
  \label{fig:velocity_correction_hdc_difference}
\end{figure}

The spectra were also corrected for their radial  and barycentric velocity. Then we normalize the spectra and convert them to velocity profiles around the central wavelength of the [OI]6300 \AA ~line (6300.304 \AA). The resulting spectra of \hda and \hdc are shown in Figures~ \ref{fig:velocity_correction_hda_difference} and \ref{fig:velocity_correction_hdc_difference} while the spectrum of \hdb in Figure \ref{fig:velocity_correction_hdb_difference} needs to undergo one more reduction step to correct for the underlying photospheric absorption lines as discussed in section \ref{sec:hdb}.

 The spectra  of \hda and \hdc both show a double peaked [OI] line profile indicating the presence of a circumstellar disk and temporal variance indicating disk inhomogeneity. \hda shows a maximum signal of 10 \% above the continuum, has 60 km s$^{-1}$ wide wings and displays next to the double peak two extra 'shoulders'. The data also show temporal change with a more pronounced double peak in the B1 data as well as more flux in the shoulders. We deduce that the difference between the two lines on the red wing at 50 km s$^{-1}$ in the B2 data seen in Figure \ref{fig:velocity_correction_hda_difference} is not due to temporal variability because it drops below the continuum. It is also not present in the [OI] emission at 6363 \AA ~and thus real absorption. This line coincides with the telluric absorption in the B1 data and therefore does not show in that spectum. We have compared this spectrum with that of HD34364, another B9.5V star but failed to identify the absorption feauture, thus ruling out photospheric absorption. We also exclude telluric emission because the width of the line of 15 km s$^{-1}$ is larger than the telluric line width of 4 km s$^{-1}$. We therefore conclude that the nature of this absorption line is probably interstellar, but we cannot identify the exact origin. For further analysis we replace the affected part of the spectrum between 35 and 65 km s$^{-1}$ with the -65 to -35 km s$^{-1}$ data of the same spectrum\footnote{ In order to compare how much this changes our final results we show the results obtained with the two seperate wings in Figure \ref{fig:HD101412_endresult_leftright} as well as the results obtained with the above mentioned changes in Figure \ref{fig:INTvsRAD_HD101412}.}. We show this replacement with the grey dotted line, while the original absorption is shown with a grey solid line in Figure \ref{fig:velocity_correction_hda_difference}.

The [OI] emission line seen in \hdc rises a maximum of 5 \% above the continuum, is 50 km s$^{-1}$ wide and also shows intrinsic temporal change. The red wing of the line changes between both data sets; This is not caused by normalizing effects.   % and a velocity distribution of  110 km s$^{-1}$ wide.

The [OI] emission in \hdb has a maximum of approximately 8 \% above the continuum and a single peaked profile. Because of the overlapping photospheric absorption lines it is difficult to accurately determine the velocity broadening of the line. We estimate it to be inbetween 60 and 100 km s$^{-1}$. The [OI] line shows no descernible temporal change and, contrary to \hda and HD179218, peaks at 3.9 km s$^{-1}$ (3 $\sigma$) away from the expected stellar photospheric zero velocity.

% An upper limit on the inclination derived by matching the breakup velocity of  538 km s$^{-1}$ to the projected rotational velocity yields an upper limit of 89.2 degrees.

\section{Stellar Parameters}
\label{sec:stel_par}

\begin{figure}
  \resizebox{\hsize}{!}{\includegraphics[angle=270]{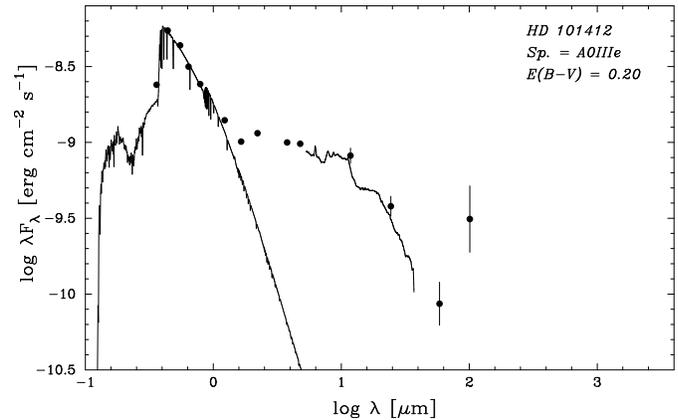}}
     \caption{The SED of \hda compiled using literature photometry. The solid line is a reddened Kurucz stellar atmosphere model fitted to the photometry of the central star. Also shown are SPITZER-IRS data from Bouwman et al. (2007, in prep.)}
     \label{fig:SED_HD101412}
\end{figure}

\begin{figure}
  \resizebox{\hsize}{!}{\includegraphics[angle=270]{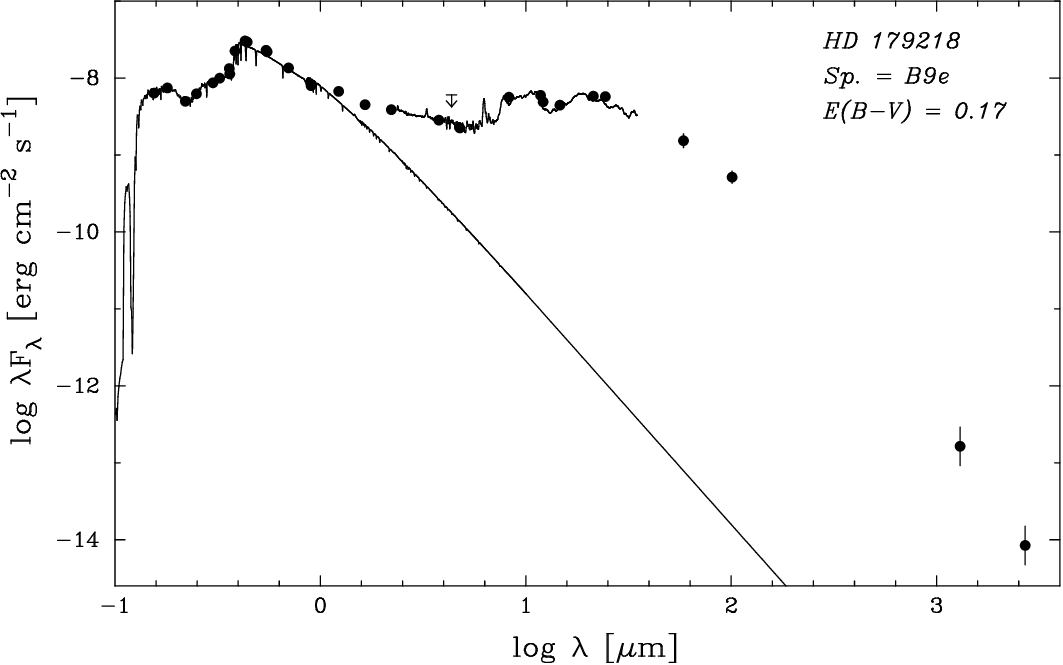}}
     \caption{Same as Figure~\ref{fig:SED_HD101412} for \hdc but the ISO-SWS spectrum is shown here \citep{2004A&A...426..151A}}
     \label{fig:SED_HD179218}
\end{figure}

\begin{figure}
  \resizebox{\hsize}{!}{\includegraphics[angle=270]{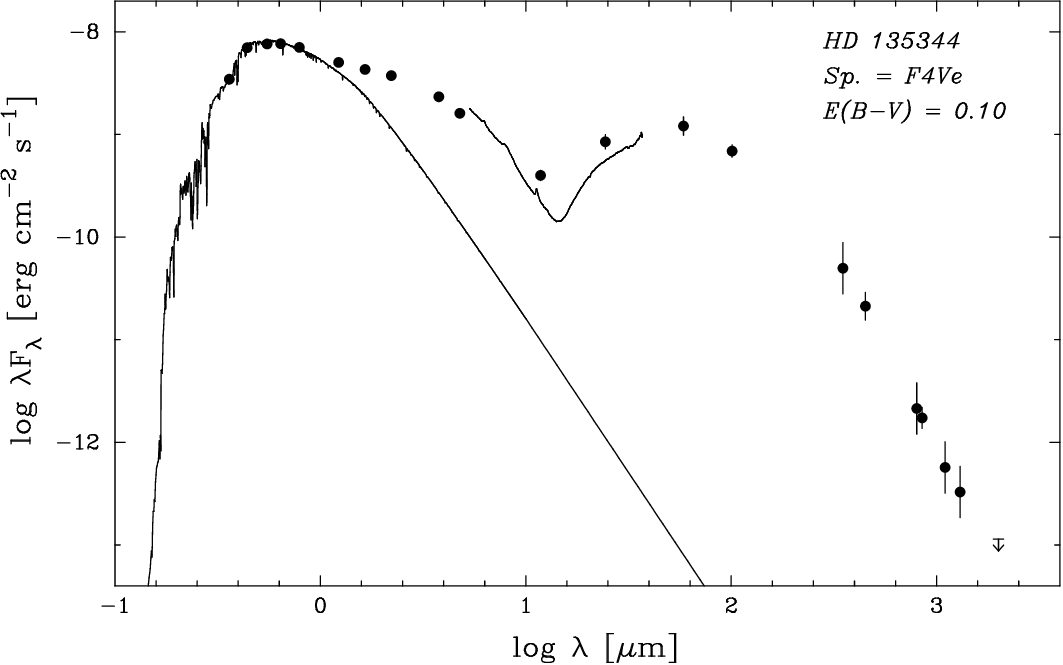}}
     \caption{Same as Figure~\ref{fig:SED_HD101412} for \hdb}
     \label{fig:SED_HD135344}
\end{figure}

The Spectral Energy Distributions (SED) of HAEBE stars can be divided into two groups based on the shape of the mid-IR (20-100 $\mu $m) emission; groups I and II \citep{2001A&A...365..476M}. Group I sources have a rising mid-IR flux excess indicative of a flared disk, while group II sources display a more modest IR excess indicating a geometrically flat disk. This division between groups is made more quantitative by a classification using the $L_{NIR}/L_{IR}$ vs. IRAS [12] - [60] colour diagram (c.f. \cite{2004A&A...422..621A}). We show the SED for each of our sources in Figures~\ref{fig:SED_HD101412} - \ref{fig:SED_HD135344}. By using this classification and comparing the SEDs, \hdc and \hdb are group I sources. In the $L_{NIR}/L_{IR}$ vs. IRAS [12] - [60] diagram \hda is located close to the line separating the two groups. Although it is a group II source, it may be considered a transitional object in between both groups. 

A \cite{1991ppag.proc...27K} model atmosphere corresponding to the photospheric parameters of the central star is fitted through the data to represent the photospherical contribution. From this, the shape and excess emission from the disk become clear. The shortest wavelength for which the excess is discernible is 0.9 $\mu $m for HD101412, 1.5 $\mu $m for \hdb and 1.0 $\mu $m for \hdc. From here on, the excess emission for \hda shows a steady decline that 'flares up' at 100 $\mu $m. This behavior is displayed by more IRAS sources at long wavelengths and is probably caused by the inclusion of interstellar material heated up by, but not directly associated with, the central star in the rather large (1 arcmin) IRAS beam. The other two SEDs show a re-brightening at mid-IR wavelengths. The fractional luminosity of the dust compared to the stellar luminosity, $L_{{\rm IRE}}/L_{\star}$, is listed in Table~\ref{table:stellar_parameters}.

To determine the projected rotational velocity, the $v\sin{i}$, we have fitted a Doppler profile to mimic the broadening of the lines by stellar rotation to the rotational line profile of several atmospheric absorption lines.%\sout{  We use these values to place a lower limit on the inclination for each of our sources by matching their theoretical breakup velocity to the $v\sin{i}$ values. This approach yields lower limits that agree with (though are less constraining than) the inclinations found in literature for \hdb and \hdc\footnote{i $\geq$ 12$^\circ$ for \hdb and i $\geq$ 9.5$^\circ$ for \hdc, see Table~\ref{table:stellar_parameters} for the literature inclination values} and only places a very weak lower limit of 0.85 degrees on the inclination of \hda.}

\begin{table*}
\caption{Astrophysical parameters of the programme stars}              % title of Table
\label{table:stellar_parameters}      % is used to refer this table in the text
\centering  
\begin{minipage}{\textwidth}                              % used for centering table
\centering                                      % used for centering table
\begin{tabular}{c c c c c c c c c }          % centered columns (4 columns)
\hline\hline                        % inserts double horizontal lines
Name & Spectral Type & $log T_{eff}$ & log $L_{bol}$ & $L_{IRE}$ & M & Distance & $v_{rad}$ & $v\sin{i}$  \\    % table heading
& & log [K] & log [$L_{\odot}$] & [$L_{\star}$] & [$M_{\sun}$] & [pc] &  [km s$^{-1}$] & [km s$^{-1}$] \\
\hline                                   % inserts single horizontal line
    \hda & A0IIIe & 4.02 & 1.40 & 0.27 & 2.3 $\pm$ 0.2 & 160\footnote{\cite{2005A&A...436..209A}}  & 16.9$ \pm$ 0.2  & 8 $\pm$ 1   \\      % inserting body of the table
    \hdb & F4Ve   & 3.82 & 1.01 & 0.43 & 1.7 $\pm$ 0.2 & 140\footnote{\cite{2005A&A...437..189V}}  & 1.6 $\pm$ 1.3 & 75 $\pm$ 5   \\
    \hdc & B9e    & 4.02 & 1.88 & 0.28 & 2.7 $\pm$ 0.3 & 240\footnote{\cite{1998A&A...330..145V}}  & 15.4 $\pm$ 2.3  & 72 $\pm$ 5    \\
%CAREFULL: DENT et al PAPER dubbel!
\hline                                             %inserts single line
\end{tabular}
%\end{minipage}
%\end{table*}

\vspace{1cm}
%\begin{table*}
%\caption{Astrophysical parameters of the programme stars, continued}              % title of Table
\centering  
%\begin{minipage}{\textwidth}                              % used for centering table
\centering                                      % used for centering table
\begin{tabular}{l l c c c c c}          % centered columns (4 columns)
\hline\hline                        % inserts double horizontal lines
Name & [OI] EW & L([OI]) & [OI] inner radius & [OI] outer radius & Inclination & Group\\    % table heading
&  [\AA] & [$L_{\odot}$] & [AU] & [AU] & [$^\circ$]& \\
\hline                                   % inserts single horizontal line
    \hda & -0.13$ \pm$ 0.01& $(9.5  \pm 0.5)\ 10^{-5}$  & 0.15 & 10 & 30  & II \\      % inserting body of the table
    \hdb & -0.13 $ \pm$ 0.06& $(1.3 \pm 0.5)\ 10^{-4} $ & 0.1 & 90 & 45\footnote{\cite{2006A&A...460..117D}}       & I       \\
    \hdc & -0.053$ \pm$ 0.005& $(3.4 \pm 0.3) \ 10^{-4}$ & 0.4 & 65 & 40\footnote{\cite{2005MNRAS.359..663D}}       & I       \\
%CAREFULL: DENT et al PAPER dubbel!
\hline                                             %inserts single line
\end{tabular}
\end{minipage}
\end{table*}

\subsection{\hda}

\hda has a spectral type between B9.5Ve \citep{1975mctd.book.....H} and A0 III/IVe  \citep{2006A&A...457..581G} and has a distance of 160 pc. By integrating under the SED shown in Figure \ref{fig:SED_HD101412} (using a reddened Kurucz model to interpolate between observed data points in the optical, a spline in the infrared and a grey body to extrapolate they last data points), and adopting this distance of 160 pc, we derive a bolometric luminosity of 25 $L_\odot$ for HD101412. The Spitzer-IRS spectrum of HD 101412 has an unusual shape due to the presence of strong PAH emission bands on top of a plateau-like 10 $\mu$m silicate feature over a weakly rising continuum (Bouwman et al. 2007, in prep.).  The same authors also note the presence of  emission bands due to the presence of crystalline silicates at 19, 24, 28 and 34 $\mu$m. This presence of strong PAH emission is, just like the presence of strong [OI] emission, unexpected for a group II source \citep{2004A&A...426..151A}. This reinforces our suggestion that the disk of \hda is in transition between flaring and being self shadowed.

 We estimate the mass of HD101412, derived by comparing the position in the HR diagram to the theoretical pre-main-sequence tracks of \cite{1993ApJ...418..414P}, to be 2.3 $\pm$ 0.2 $M_\odot$.  \cite{2006A&A...457..581G} find  a v $\sin{i}$ of 7 $\pm$ 1 km s$^{-1}$ by modeling high resolution (R = 48000) spectra, suggesting we see \hda almost pole-on. This agrees well with the value derived from our data of 8 $\pm$ 1 km s$^{-1}$. %\sout{A lower limit on the inclination of 0.85 degrees was derived by matching the breakup velocity of  538 km s$^{-1}$ to the projected rotational velocity.} %In this paper we use a lower than averagely expected value for the inclination of 30 degrees, because both the low  v $\sin{i}$ and the prominent double peak suggest a small inclination angle.

\subsection{\hdb}

\hdb is a wide visual binary with a separation of 19.6 arcsec located at a distance of 140 pc. In this paper we will discuss component \hdb which has spectral type F4Ve and an estimated mass of 1.7 $\pm$ 0.2 $M_\odot$. It has a bolometric luminosity of 10 $L_\odot$, obtained by fitting the SED shown in Figure \ref{fig:SED_HD135344}. Bouwman et al. (2007, in prep.) note only very weak PAH features in the Spitzer-IRS spectrum of HD135344 B.  They also note that its SED is quite unusual, with an IR excess starting at 1.5 $\mu$m, but a deep bump at 13 $\mu$m. \cite{2006A&A...460..117D} have imaged \hdb in the mid-infrared (20.5 $\mu$m) and find a disk with an inclination of 45$^\circ$ and a PA of 100 $\pm 10^\circ$. Based on these spatially resolved observations they estimate the disk to extend up to 200 AU. %\sout{If we compare the breakup velocity of 364 km s$^{-1}$ with the $v\sin{i}$  the inclination  is constrained to i $\geq$ 12$^\circ$.}
 \cite{2007ApJ...664L.107B} classify the disk around \hdb as a 'cold disk', interpreting the lack of mid-IR emission in the SED as missing warm dust. From modeling the SED they derive the disk to have an inner rim starting at 0.18 AU, a gap (i.e. missing warm dust) from 0.45 to 45 AU and outside that a gas rich disk. 

A simple model fit to the CO(J=3-2) double peaked line profile done by \cite{2005MNRAS.359..663D} indicates an inner radius of the emitting gas $\leq 10$ AU, an outer radius of 75 $\pm$ 5 AU, an estimated circumstellar dust mass of $10^{-4}M_\odot$ and an inclination of 11$\pm 2^\circ$.  The value of this inclination is notably different from the result of \cite{2006A&A...460..117D} and implies that the star rotates at breakup velocity. We will use the result of \cite{2006A&A...460..117D} in our model because it is obtained in a more direct way. 

%The radial velocity as determined from Figure~\ref{fig:radial_velocity_variation_HD135344} is 1.7 km s$^{-1}$. If we apply this correction to the spectra the resulting line profile is not symmetric around 0 km s$^{-1}$  but appears to be shifted with a couple of km s$^{-1}$. Since we expect it to be symmetric we have  used another radial velocity (4.5 in stead of 1.6 km s$^{-1}$) for the modeling. The plot shown in Figure~\ref{fig:velocity_correction_hdb} is made using this 'corrected'  radial velocity. 

\subsection{\hdc}

\hdc is a B9e star at a distance of 240 pc and is the Herbig star with the $2^{nd}$ largest known percentage of crystalline dust \citep{2005A&A...437..189V}. It has a bolometric luminosity of 76 $L_\odot$, obtained with the SED shown in Figure \ref{fig:SED_HD179218}. HD179218 is the star with the greatest richness in terms of number of infrared spectral features in the overview of infrared spectra of Herbig Ae/Be stars by Bouwman et al. (2007, in prep.).  Its Spitzer-IRS spectrum is dominated by strong emission features due to the presence of abundant crystalline silicates.  Polycyclic Aromatic Hydrocarbons are also detected, in agreement with its classification as a group I source \citep{2004A&A...426..151A}. We estimate the mass based on the position in the HR diagram to be 2.7  $\pm$ 0.3 $M_\odot$.  \cite{2006A&A...457..581G} find a v $\sin{i}$ of 72  $\pm$ 3 km s$^{-1}$ agreeing well with the value derived from our data of 72 $\pm$ 5 km s$^{-1}$.  

\cite{2004ApJ...601.1000E} fit the continuum spectrum of \hdc with a model including a polar cavity, suggesting a disk-like geometry in the innermost part of the envelope. Moreover, it is noteworthy that their model  corresponds to a dust absorption coefficient exponent $\beta = 0.6$ which is typical of large dust grains and agrees with the results of \cite{2004A&A...422..621A}. This is in agreement with the idea that the grain size is expected to increase with time in circumstellar envelopes, as a result of the ongoing growth processes activated in high-density environments. 
\cite{2007ApJ...658.1164L} have resolved warm dust using 10 micron nulling interferometric observations around the star with a diameter of 27 $\pm$ 5 AU. Their observations suggest circular symmetry and thus a small inclination, but inclinations up to 45$^\circ$ are still within the error bars.
A simple model fit to the CO(J=3-2) line profile done by \cite{2005MNRAS.359..663D} indicates an outer radius of 120 $\pm$ 20 AU, an inclination of 40$\pm 10^\circ$ and an estimated circumstellar dust mass of $10^{-4}M_\odot$.

\section{Data Analysis}
\label{sec:analysis}

\subsection{Method }
\label{sec-method}
In pioneering work on the mechanism responsible for [OI] emission in young stars, \cite{1995ApJ...452..736H} modeled the broad [OI] 6300 \AA\, line emission seen in T Tauri stars as being due to a combination of dense stellar jets and a disk wind or magnetic accretion columns. \cite{2005A&A...436..209A}  have developed a model that explains the narrow [OI] emission lines seen around many HAEBE stars.%\sout{ model the [OI] emitting region of HAEBE stars to translate the Doppler broadened emission into an intensity versus radius distribution. They assume in their model that the [OI] emission is non-thermal and comes from the photo dissociation of OH and H$_{\rm{2}}$O molecules in the surface layers of the disk by the UV radiation field of the central star and a passive flared disk in keplerian rotation.}
  This simplified model assumes that the entire UV luminosity (2-13.6 eV) of the central star (represented by a Kurucz model for the stellar photosphere) is used to photo dissociate OH in the upper layers of the disk. The disk structure (i.e. densities and temperatures) is computed with a \cite{1997ApJ...490..368C} type disk, with small olivine dust grains as the opacity source. The OH to H abundance is fixed a priori and constant at each location in the disk. The absorption of UV photons by the dust in the disk is taken into account in the same way as in the models of \cite{2000ApJ...539..751S}  for photo dissociation regions along the direction of the in falling photons. The line profiles are deduced assuming Keplerian rotation and integration over the spatial coordinates. Details about the method we employ here can be found in  \cite{2005A&A...436..209A} and \cite{2006A&A...449..267A}.

%10. There is a rather long description of why one rejects a thermal origin for both 5577 and 6300. This should be collapsed to a single sentence that states that if the temperature of the gas is like that of the disk (i.e., ~ 1000K) then the Boltzmann factor renders the 5577/6300 ratio very low. Of course, thermal origins are still possible if the disk has a hot corona or wind. In fact, section 4.2.1 indicates that the blue side of the line profile has more flux than the red side, as one would expect if there is a component of a disk wind to the line profile, but no mention is made of this possibility. 

\begin{figure}
  \resizebox{\hsize}{!}{\includegraphics{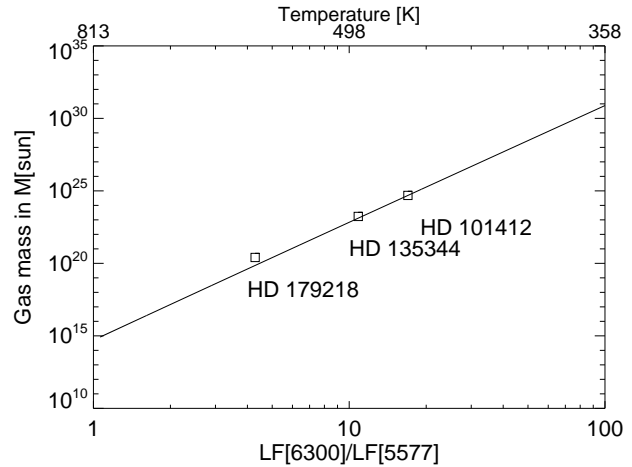}}
     \caption{Total mass of gas needed to explain the received [OI] 6300 \AA\, flux vs. the relative strength of the 5577 \AA\ vs. 6300 \AA\ line (bottom) in case of thermal emission. The corresponding temperature that is needed to explain this ratio is plotted on the upper x-axis. Also shown are the data points for the 3 stars (squares) and the function connecting the ratio of Equivalent Widths to the total gas mass for \hda (solid line).  Because both lines are only resolved for \hda we show lower limits for the other two targets. The disks around our three targets would need to be unrealistically massive to explain the observed [OI] emission as being due to thermal emission from the disk}
     \label{fig:lineratio}
\end{figure}

 To verify the validity of this assumption for the three systems we study here we compare the Line Flux (LF) of the 5577 \AA\, [OI] emission to that of the 6300 \AA\, [OI] emission. If this emission is caused by thermal emission originating from a disk with a temperature of 500 K, the Boltzmann factor renders the ratio of these two very low (Figure \ref{fig:lineratio}).%\sout{In case of a thermal emission mechanism the ratio of the two lines translates directly to a temperature - assuming a Boltzmann distribution for the oxygen atoms and an absence of other collisional excitations caused by e.g. hydrogen atoms or free electrons - for the emitting gas. This ratio is a simple and maximum approximation for the ratio of the EW's given a certain temperature. If we use the temperature corresponding to the EW ratio (between 450 and 600 K) and calculate how much gas of this temperature is needed to explain the received flux, it becomes clear that the emission mechanism cannot be thermal.} 

%\sout{Using our stellar parameters we expect that the EW ratio of $\frac{EW(6300)}{EW(5577)}$ for thermal emission is $\leq$ 0.03.
% Both emission lines are only resolved for \hda, giving a ratio  of 19. Because the 5577 \AA\, [OI] emission is not detected in the other 2 spectra, we derive lower limits for the ratio by assuming that the FWHM of both lines is the same, and by taking a 3$\sigma$ detection limit. These lower limits are 10 for \hdb and 4 for HD179218 and are plotted in Figure \ref{fig:lineratio}. We conclude that the disk around our 3 targets would need to be unrealistically massive to explain the observed [OI] emission as being due to thermal emission from the disk. Instead,} 
We follow \cite{2005A&A...436..209A} and \cite{2006A&A...449..267A} in concluding that the observed [OI] emission is caused by illumination of the upper disk layer by UV radiation. 

%\begin{figure}
%  \resizebox{\hsize}{!}{\includegraphics{lineratio.eps}}
%     \caption{The relative strength of the 5577 \AA\ vs. 6300 \AA\ line. Also shown are the data points for the 3 stars and the distance from the central star matching the corresponding temperature}
%     \label{fig:lineratio}
%\end{figure}

%\begin{figure}
%  \resizebox{\hsize}{!}{\includegraphics{lineratio_fraction.eps}}
%     \caption{Total mass of circumstellar disk needed to explain the received [OI] 6300 \AA\, flux in case of a thermal emission mechanism in solar masses}
%     \label{fig:lineratio_fraction}
%\end{figure}

  We expect that the pattern and intensity of the [OI] emission is influenced by differences in disk geometry. The part of the disk that lies in the shadow of the puffed up inner rim characteristic for group II sources will emit less compared to the parts that are directly illuminated by the central star. Therefore we predict that there is a drop in intensity at the location of the shadow cast by the inner rim in HD101412. How extended this shadow is depends on when -and if- the disk rises out of the shadow again. We also expect the flaring disks to appear larger in [OI] emission, because they are illuminated by stellar UV flux at greater distance and with more intensity due to their shape than the self shadowing disks.

We note that the radius scale of our results is dependent on the inclination of the disk used in the model. Taking a smaller inclination angle shifts the emission region closer to the star. If we compare the onset of the [OI] emission with the location of the dust sublimation radius, we see that these agree for the inclinations used for \hdb and \hdc. We use this to guess an inclination for HD101412. Values between 24$^{\circ}$ and 41$^{\circ}$  allow the inner rim to be at the dust sublimation radius. In the rest of this paper we will use a value of 30$^{\circ}$. This low inclination is in agreement with both the low v $\sin{i}$ and the double peaked line profile seen in [OI].

\subsection{Results}

\subsubsection{\hda and \hdc}
\begin{figure}
  \resizebox{\hsize}{!}{\includegraphics{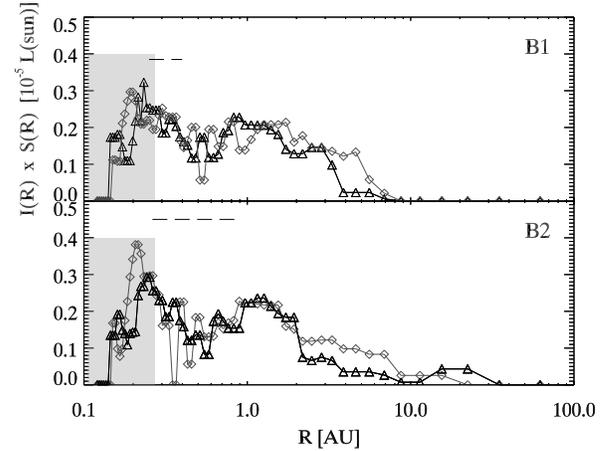}}
     \caption{The [OI] emission of the blue shifted (grey, diamonds) and red shifted (black, triangles) halves of the disk of \hda. The two panels show the data taken at different times. The grey area is the suspected location of the inner rim. The horizontal dashed lines mark in the upper plot the location of the removed telluric absorption line (Figure \ref{fig:telluric_correction_101412}), and in the lower plot the part of the B2 spectrum that has been replaced as described in section \ref{sec:obs} and Figure \ref{fig:velocity_correction_hda_difference}}
     \label{fig:HD101412_endresult_leftright}
\end{figure}

\begin{figure}
  \resizebox{\hsize}{!}{\includegraphics{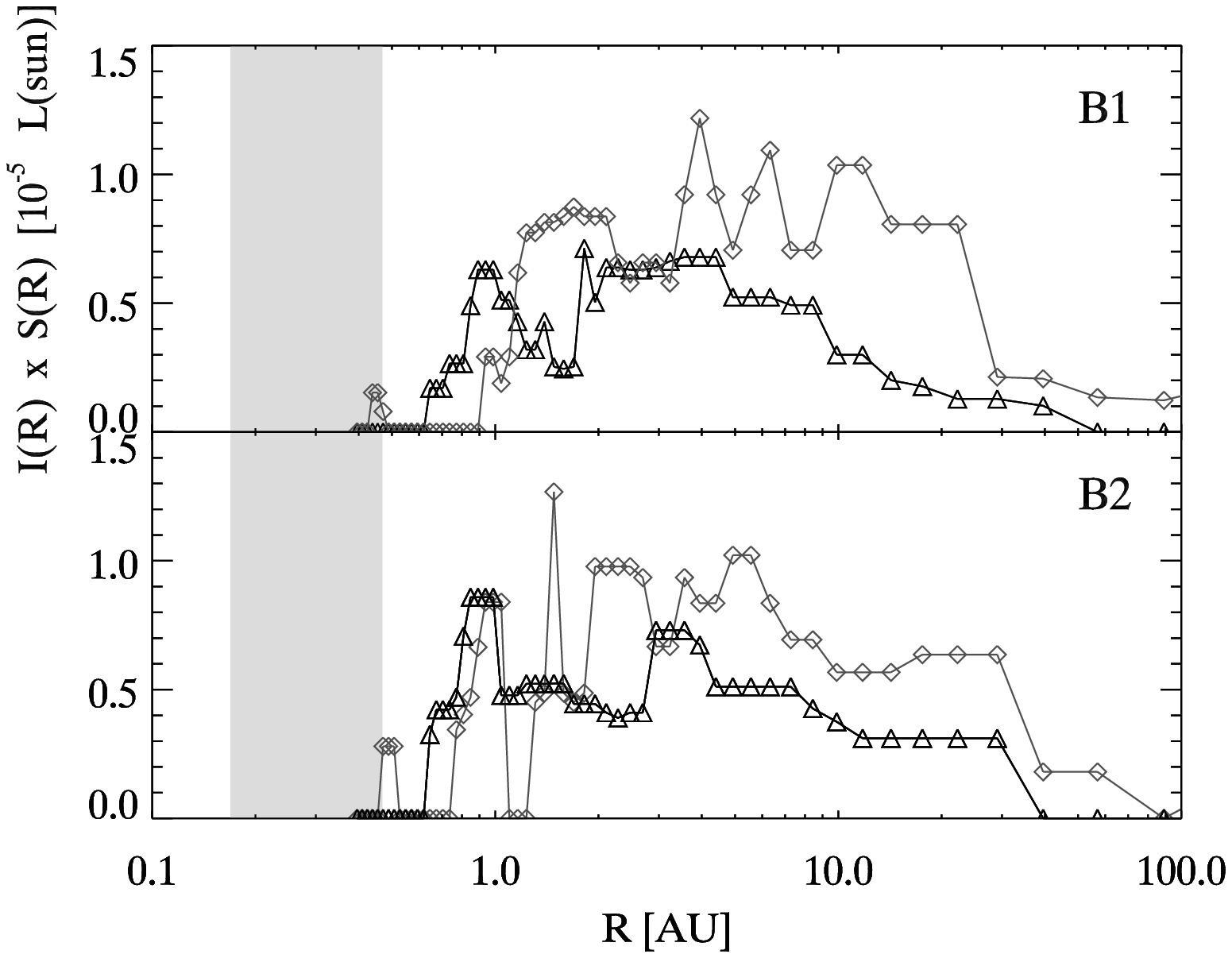}}
     \caption{Same as Figure \ref{fig:HD101412_endresult_leftright} for \hdc}
     \label{fig:HD179218_endresult_leftright}
\end{figure}

By comparing the blue- and red shifted Doppler profiles, it is possible to distinguish between the parts of the disk that rotate towards us or away and thus track possible inhomogeneities. We have done this for \hda and \hdc and show the results in Figures \ref{fig:HD101412_endresult_leftright} and \ref{fig:HD179218_endresult_leftright} plotted together with the dust sublimation radius for temperatures between 1200 and 2000 K in grey. The blue- and red-shifted wings of \hda look approximately the same in the upper (B1) and bottom (B2) graph apart from some excess flux at larger distances in the blue shifted part of the disk. \hdc looks mildly asymmetric, showing for the red shifted part emission that starts at a closer distance to the star and lacking intensity at larger distances compared to the blue shifted part. The blue side also seems to harbor more flux and is detectable up to larger distances. The results discussed below for \hda and \hdc are obtained by averaging the red- and blue shifted profiles of both stars.

\begin{figure}
  \resizebox{\hsize}{!}{\includegraphics{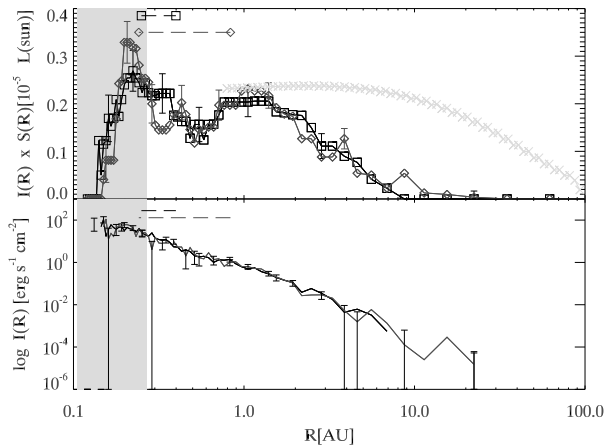}}
     \caption{The intensity versus radius graph of \hda showing the distribution of [OI] emission as a function of distance from the central star. The two lines represent data taken on different dates; B1 is plotted in black with squares and B2 in grey with diamonds. The grey area is the suspected location of the inner rim and the light grey crossed line represents a simple model for [OI] emission from a flared disk in Keplerian rotation. In the {\it upper} plot we show the intensity normalized to the disk surface in a semi-log plot, as to better compare the inner and outer parts of the disk. A drop in intensity can clearly be seen at 0.6 AU from the central star. In the lower plot we show the intensity derived from our model on a log - log scale. The grey horizontal dashed lines marks the part of the red shifted part of the B2 spectrum that has been replaced as described in section \ref{sec:obs} and the black horizontal dashed lines marks the part of the red shifted part of the B1 spectrum of the removed telluric absorption line.}
     \label{fig:INTvsRAD_HD101412}
\end{figure}

\begin{figure}
  \resizebox{\hsize}{!}{\includegraphics{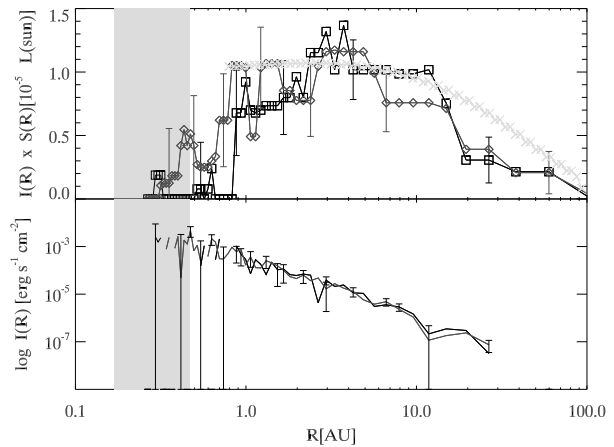}}
     \caption{Same as Figure~\ref{fig:INTvsRAD_HD101412} for \hdc. The spikes at small radii and at 4 AU are an artifact of our data reduction and not significant.}
     \label{fig:INTvsRAD_HD179218}
\end{figure}

The distribution of emitting atomic oxygen gas as a function of radius for HD101412 is plotted in Figure~\ref{fig:INTvsRAD_HD101412} and for HD179218 in Figure~\ref{fig:INTvsRAD_HD179218}. We have plotted the log - log dependency of the received flux as well as a semi log scale plot where we have normalized the intensity with the surface of the ring at distance R : S(R) where $S(R)$ = $\pi* (R_{out}^2 - R_{in}^2)$ with $R_{out}$ and $R_{in}$  the outer and inner radius of the ring at distance R. In this sense $I(R)xS(R)$ is the total luminosity in a ring at distance R and the fluxes of the inner and outer part of the disk can be directly compared this way. The error bars drawn in these graphs are solely a function of the signal to noise of the data.  The [OI] emission of \hda starts within the expected location of the inner rim at 0.15 AU and extends to 10 AU with a clear drop of almost 50 $\%$ in the I(R) vs. S(R) curve around 0.5 AU. Since \hda is a group II source, characterized by a puffed up inner rim, and because the [OI] emission mechanism is directly proportional to (stellar) UV flux, the drop in intensity can be caused by the shadowing effect of the inner rim. Once the disk emerges out of the shadow (around 1 AU), the disk is again directly illuminated by the central star and keeps receiving enough flux up to 10 AU to be detectable. If we compare the [OI] emission of \hda with that expected from a flared disk in Keplerian rotation, shown with light grey crosses in Figure \ref{fig:INTvsRAD_HD101412}, one sees that for a flared disk the emission should have been detected up to much larger radii. We interpret these discrepancies with the flared disk model that both can be explained by a self-shadowed disk; the local drop in intensity at 0.6 AU and the dissapearance of emission at 10 AU, as direct evidence  that in the disk surrounding HD101412, the inner rim indeed shadows part of the outer disk - as expected for a group II source.

The intensity vs. radius distribution for HD179218, a group I source, looks different. It starts at 0.3 AU from the central star, extends\footnote{ This number given here is larger than the in section \ref{sec:obs} inferred upper limit for the disk size. This is caused by the fact that the disk emission is concentrated towards the center and drops of exponentially as can be seen e.g. in the log - log part of Figure \ref{fig:INTvsRAD_HD101412}. The the emission originating from 2 AU of the central star already has dropped in intensity by a factor of 100 compared to the maximum.} up to 65 AU and has no clearly discernible local drop in intensity as seen in HD101412.  This disk has a flaring geometry, meaning that it has - depending on the disk models used - no or negligible shadowing by the inner rim.  The flaring also allows the disk to intercept more stellar flux at larger distance, causing the [OI] emission up to a larger distance from the central star. The onset of emission for the B2 data of \hdc is located beyond the dust sublimation radius. This result is reached by assuming the inclination found by   \cite{2005MNRAS.359..663D} of  40$\pm 10^\circ$. When the lower limit of the inclination of $30^\circ$ is assumed, the onset of emission starts at the dust sublimation radius. Both the shape and the inner plus outer radii of the [OI] emission are in reasonable agreement with those expected from a flared disk model. The discrepancy between the observed and modeled emission at large radii is also seen in the two group I sources investigated by \cite{2006A&A...449..267A}. This may be due to the model assumption that the OH abundance is homogeneous throughout the disk.

%;\begin{figure}
%;  \resizebox{\hsize}{!}{\includegraphics{result_scales_HD101412_B1.eps}}
%;     \caption{The intensity versus radius graph of \hda showing the distribution of [OI] emission as a function of distance from the centr;al star for the data taken during the B1 period. The top graph is normalized on the surface of the ring, so that the luminosity close to t;he star can be directly compared to that further away. The bottom graph shows the log log plot of the Intensity vs. radius.}
%;     \label{fig:result_scales_HD101412_B1}
%;\end{figure}

%;\begin{figure}
%;  \resizebox{\hsize}{!}{\includegraphics{result_scales_HD101412_B2.eps}}
%;     \caption{The intensity versus radius graph of \hda showing the distribution of [OI] emission as a function of distance from the centr;al star for the data taken during the B2 period. The top graph is normalized on the surface of the ring, so that the luminosity close to t;he star can be directly compared to that further away. The bottom graph shows the log log plot of the Intensity vs. radius.}
%;     \label{fig:result_scales_HD101412_B2}
%;\end{figure}

%\begin{figure}
%  \resizebox{\hsize}{!}{\includegraphics{result_scales_HD179218_B1.eps}}
%     \caption{The intensity versus radius graph of \hdc showing the distribution of [OI] emission as a function of distance from the central star for the data taken during the B1 period. The top graph is normalized on the surface of the ring, so that the luminosity close to the star can be directly compared to that further away. The bottom graph shows the log log plot of the Intensity vs. radius.}
%     \label{fig:result_scales_HD179218_B1}
%\end{figure}

%\begin{figure}
%  \resizebox{\hsize}{!}{\includegraphics{result_scales_HD179218_B2.eps}}
%     \caption{The intensity versus radius graph of \hdc showing the distribution of [OI] emission as a function of distance from the central star for the data taken during the B2 period. The top graph is normalized on the surface of the ring, so that the luminosity close to the star can be directly compared to that further away. The bottom graph shows the log log plot of the Intensity vs. radius.}
%     \label{fig:result_scales_HD179218_B2}
%\end{figure}

\subsubsection{\hdb}
\label{sec:hdb}

\begin{figure}
  \resizebox{\hsize}{!}{\includegraphics{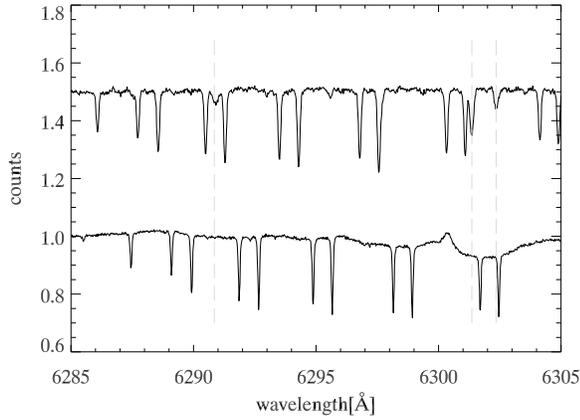}}
  \caption{The spectra of \hdb (bottom) and HIP53070 (top), another F4V star but with low v $\sin{i}$. This plot shows that there are two broadened absorption lines marked with vertical lines mixing with the [OI] emission line of \hdb. All the other absorption lines are caused by telluric absorption.}
  \label{fig:HIP_and_voigt}
\end{figure}

\begin{figure}
  \resizebox{\hsize}{!}{\includegraphics{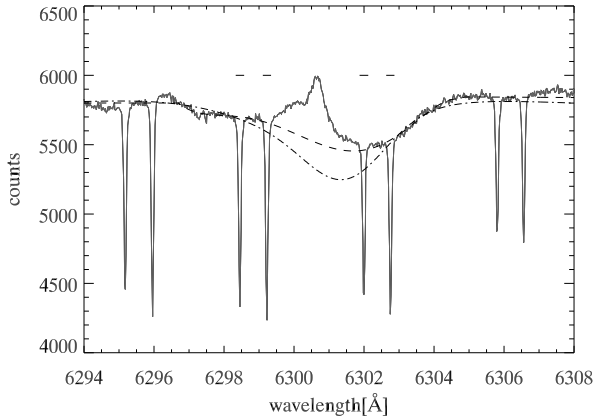}}
  \caption{The Chebyshev (dashed line) and the double Voigt (dot-dashed line) model used to model the photospheric absorption in \hdb. The telluric absorption lines at the location of the four horizontal lines have been clipped out.}
  \label{fig:cont_models}
\end{figure}

\begin{figure}
  \resizebox{\hsize}{!}{\includegraphics{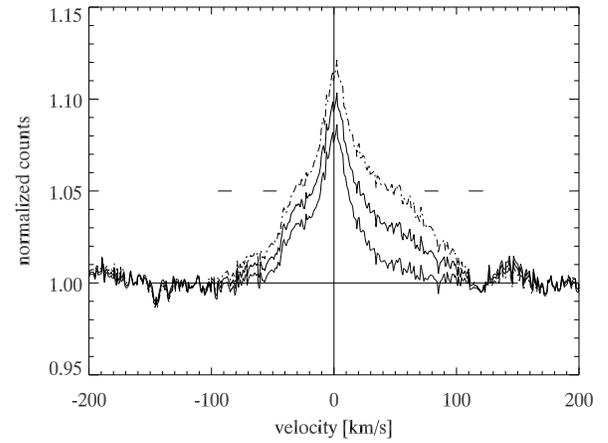}}
  \caption{The velocity corrected, normalized and converted to velocity around the [OI] line spectra corrected for the photospheric absorption using the 3 models of \hdb. From top to bottom: the double Voigt model, the average model and the Chebyshev model. The horizontal bars mark the location of the clipped out telluric absorption lines. Only the spectra of B1 are shown}
  \label{fig:velocity_correction_hdb}
\end{figure}

\begin{figure}
  \resizebox{\hsize}{!}{\includegraphics{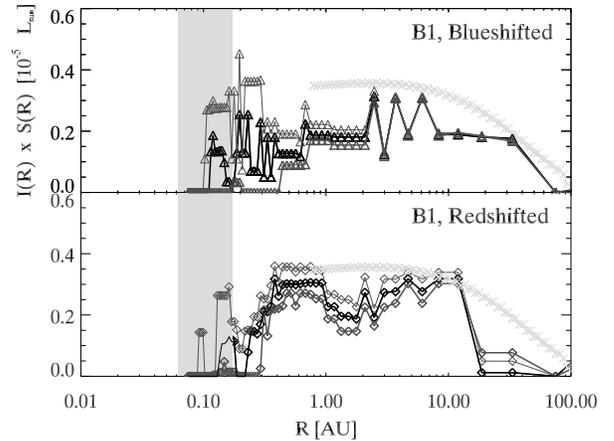}}
     \caption{The intensity versus radius graph for the blue (upper graph, triangles)- and red(lower graph, diamonds) shifted sides of \hdb showing the distribution of [OI] emission as a function of distance from the central star. The three lines represent the B1 measurements for the Voigt model (upper grey line), the Chebyshev model (lower grey line) and the average of the two (thick black line). the light grey smooth line represents a simple model for the [OI] emission from a disk in Keplerian rotation.}
     \label{fig:INTvsRAD_HD135344_all_leftright_B1}
\end{figure}

\begin{figure}
  \resizebox{\hsize}{!}{\includegraphics{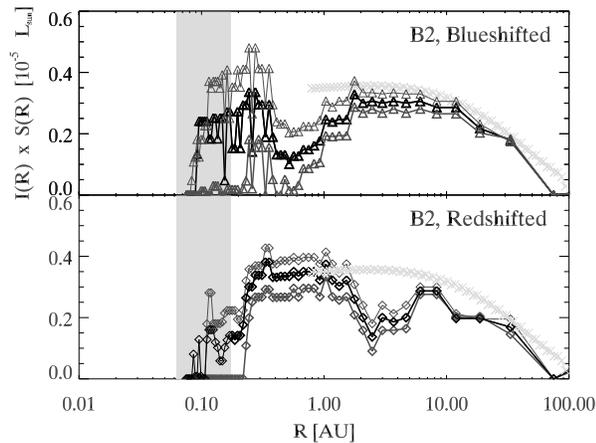}}
     \caption{Same as Figure \ref{fig:INTvsRAD_HD135344_all_leftright_B1} for the B2 measurements}
     \label{fig:INTvsRAD_HD135344_all_leftright_B2}
\end{figure}

The spectrum of \hdb is contaminated by very broad atmospheric absorption lines. Because of this and because of the fact that the [OI] feature is overlapped by two of these lines as shown in Figure~\ref{fig:HIP_and_voigt}, we have undertaken one extra step to reduce these data and therefore will discuss this source separately. We correct the absorption by these two lines with two models: (1) A double Voigt model, because a Voigt profile closely mimics the expected profile of a stellar atmospheric line and (2) a polynomial model (a good fit was found using a 6th order Chebyshev-polynomial), both shown in Figure~\ref{fig:cont_models}.
A problem when fitting the Voigt profile was that we were unable to fit the absorption using the location of the lines found in HIP53070, a reference star with  the same spectral type as \hdb (Figure~\ref{fig:HIP_and_voigt}). Using these values results in over(under) fitting the right(left) wing of the absorption. For the further modeling of \hdb we use these two models and their average, as shown in  Figure~\ref{fig:velocity_correction_hdb}. This Figure also shows that the blue- and red shifted sides are asymmetric. This can either be caused by an asymmetry in the emitting region, or by an imprecise correction for the photospheric absorption (close inspection of Figure~\ref{fig:HIP_and_voigt} suggests that the blue side of the line lies on a monotonous rising slope, while the red side of the line may overlap with the minimum of the mixed absorption line). Because of this, the blue- and red shifted sides are considered separately in Figures \ref{fig:INTvsRAD_HD135344_all_leftright_B1} and \ref{fig:INTvsRAD_HD135344_all_leftright_B2}. There is a big difference between the red and blue wing in the first AU, as well as between the different models and data procured at different times. The blue shifted profile shows a steep initial rise followed by a big drop resembling HD101412, whereas the red side rises more slowly, and displays a smaller drop in intensity at a larger radius. There is also a big difference between the two models (and thus also their average) used in the first couple of AU. For the Chebyshev model the emission consistently starts at larger radii than that of the Voigt model. In all cases however, the emission extends up to a larger distance and with more brightness than the other two sources and vanishes around 90 AU. This value is comparable with the outer radius of CO gas of 75 $\pm$ 5 AU from  \cite{2005MNRAS.359..663D}.

 %Because the data taken on different dates show little difference, we have averaged these data.

Considering (1) that it is not possible to properly correct for the photospheric absorption, (2) the fact that \hdb has an SED that is uncharacteristically for either a flaring or a self-shadowing disk because of the high (near) infrared excess between 1 and 4 $\mu$m and (3) the fact the peak of the [OI] emission velocity distribution does not coincide with the stellar photosphere we consider it likely that the observed [OI] emission could also be influenced by other components than solely a disk such as an outflow with thermally exited oxygen. We will therefore limit our interpretation of our [OI] data on this star to a purely phenomenological description of the spectra.

%\begin{figure}
%  \resizebox{\hsize}{!}{\includegraphics{radial_velocity_variation_HD135344.eps}}
%  \caption{Radial velocity determination for \hdb from 62 lines. The values for the radial velocity have a mean value of  1.7 $\pm$ 1.3 km s$^{-1}$ }
%  \label{fig:radial_velocity_variation_HD135344}
%\end{figure}

%\subsection{\hdb}
%
%Say something about \hdb, despite the (obvious) flaws in our reduction mechanisms?

%\begin{figure}
%  \resizebox{\hsize}{!}{\includegraphics{HD135344_all_endresult.eps}}
%     \caption{The intensity versus radius graph of \hdb showing the distribution of [OI] emission as a function of distance from the central star. The 3 lines represent the resulting models using respectively (top to bottom) the Voigt, average or Polynomial correction. The data for both B1 and B2 have been averaged together}
%     \label{fig:INTvsRAD_HD101412}
%\end{figure}

%\begin{figure}
%  \resizebox{\hsize}{!}{\includegraphics{HD135344_all_endresult_B1.eps}}
%     \caption{The intensity versus radius graph of \hdb showing the distribution of [OI] emission as a function of distance from the central star. The 3 lines represent the resulting models using respectively (top to bottom) the Voigt, average or Polynomial correction for the B1 data}
%     \label{fig:INTvsRAD_HD101412}
%\end{figure}

%\begin{figure}
%  \resizebox{\hsize}{!}{\includegraphics{HD135344_all_endresult_B2.eps}}
%     \caption{The intensity versus radius graph of \hdb showing the distribution of [OI] emission as a function of distance from the central star. The 3 lines represent the resulting models using respectively (top to bottom) the Voigt, average or Polynomial correction for the B2 data}
%     \label{fig:INTvsRAD_HD101412}
%\end{figure}

\section{Discussion and Conclusions}
\label{sec:dis_con}

We have detected non-thermal [OI] emission in all three targets, one of which shows a double peaked emission profile and another one showing a a tentative double peak but certainly flattened emission profile indicating emission from a circumstellar disk. We see in the [OI] emission of one of these sources, HD101412, evidence for the existence of a puffed up inner rim followed by a shadow, supporting the disk model of \cite{2001ApJ...560..957D}. 
 Although this model can explain most of our observations there is one unexpected feature; the re-brightening (the 2$^{nd}$ bump) of the disk of \hda shortly after the shadow cast by the inner rim. We expect the disk to remain in the shadow much longer and possibly never emerging from it, thus preventing the disk to become more bright. We offer a possible explanation for this behavior by speculating that the dust could have a lower scale height than the gas, allowing for a layer of gas \textit{above} the dust disk. This layer may well (partially) rise above the shadow of the inner rim causing the behavior seen in Figure \ref{fig:INTvsRAD_HD101412}. The [OI] emission of \hdc is in good agreement with emission from the surface of a flaring circumstellar disk inclined at a 30 degree angle. Finally, we encounter several difficulties in explaining the [OI] emission of \hdb by a disk and therefore suggest the possibility of more components responsible for the emission than just the disk.  

%Although our observations are mostly explained by the above mentioned disk model, there is one unexpected aspect.  presence of a visible 'shadow' in the self-shadowed and flaring disks. Following models of of e.g. \cite{2001ApJ...560..957D}. the self shadowing disk does not emerge out of the shadow of the inner rim, and thus we don't expect the re-brightening seen in \hda. For the flaring disks we do expect to see a temporary dip in intensity caused by the shadow that disappears when the disk flares out of the shadow. These things are not detected. For the group I source we see  the emission drop behind the inner rim as expected, but after a while there is an unexpected rise in intensity. This dip is not at al detected for the flaring disks. This can be caused by the fact that the inner rim rises out of the shadow sufficiently fast so that it is un-resolvable by our resolution (this can be tested by having some actual disk models, keep in mind the flexibility of the location of the drop caused by the uncertainty of the inclination). Another explanation, also explaining the behavior of \hda,  is that the dust and gas are decoupled in the disks. In this case the dust has settled more than the gas, allowing for a layer of gas \textit{above} the dust disk. This layer may well (partially) rise above the shadow of the inner rim  

We will improve on the modeling done in this paper in the next papers, where we will present MIDI observations of these stars showing warm dust enabling us to compare the coupling between gas and dust. We will then also model all our observations together with the SEDs to further test and improve on the method employed in this paper, as well as to constrain as many stellar parameters on our targets as possible.

\acknowledgements{The authors of this paper wish to thank J. Liske for his assistance with modeling the  atmospheric absorption lines.
 We also want to thank the referee for his helpful suggestions that helped to improve the presentation of the results in this paper.}
\clearpage

\bibliographystyle{aa} % style aa.bst
\bibliography{articles} % your references Yourfile.bib

\end{document}